# Frustration phenomenon in the spin-1/2 Ising-Heisenberg planar model of inter-connected trigonal bipyramid structures

Lucia Gálisová

Institute of Manufacturing Management, Faculty of Manufacturing Technologies with a seat in Prešov, Technical University of Košice, Bayerova 1, 080 01 Prešov, Slovak Republic

E-mail: `galisova.lucia@gmail.com`

**Abstract.** Ground-state and finite-temperature properties of the exactly solvable mixed spin-1/2 Ising-Heisenberg planar model composed of identical trigonal bipyramids that are arranged into a regular archimedean lattice are examined with the aim to clarify the frustration phenomenon at zero and finite temperatures. It is shown that the ground-state spin frustration persists even far above the second-order phase transition. If the interaction ratio between the Heisenberg and Ising exchange interactions is close enough to the ground-state boundaries between the neighboring phases, a remarkable re-entrance of the (non-)frustrated spin arrangement of the Heisenberg spins can be observed around the critical temperature of the model. It is also evidenced that entropy and specific heat show pronounced temperature variations not only around the critical temperature, but also in low-temperature regime if values of the interaction parameters are taken from neighborhood of the ground-state phase transitions, where energies of the neighboring phases are very close.

## 1. Introduction

Understanding magnetism of two-dimensional (2D) frustrated Heisenberg spin models belongs to central goals of modern statistical physics, as they exhibit a rich variety of fascinating physical phenomena [1–4]. Besides an academic interest, theoretical studies are often motivated by an existence of real frustrated quasi-2D magnetic compounds, such as $Cs_2CuBr_4$ [5] and $RbFe(MoO_4)_2$ [6], which have a structure of the frustrated triangular lattice, $SrCu_2(BO_3)_2$ [7] and $RB_4$ (R = Er, Tm) [8], which are well realizations of the quantum Shastry-Sutherland lattice, $[Cu_3(titmb)_2(CH_3COO)_6]\cdot H_2O$ [9], $KFe_3(OH)_6(SO_4)_2$ [10], which represents a perfect Kagomé lattice antiferromagnet, and/or $Cu_9X_2(cpa)_6$ (X = Cl, Br; cpa = carboxypentonic acid) [11], whose crystal structure can be described by the isotropic Heisenberg triangulated Kagomé lattice.

However, the investigation of frustration effects in pure Heisenberg models represents a rather complex theoretical task, often accompanied by various computational difficulties. These considerably restrict a manipulation with investigated systems, but also motivate scientists to look for novel alternative computational approaches and lattice-statistical models, which could provide the best possible understanding of experimetally observed cooperative phenomena. Naturally, rigorously (exactly) solvable models are most requested, because they provide the results that are not affected by any approximations.

Recent studies show that the mixed-spin Ising-Heisenberg models defined on decorated planar lattices, containing triangular structures [12–18], represent suitable candidates for a complex rigorous examination of the frustration phenomenon in 2D. These simplified quantum-classical (hybrid) spin systems can be exactly treated by means of the exact generalized algebraic mapping transformations [19, 20], because they admit just local quantum spin fluctuations at decorating lattice sites, while nodal lattice sites are occupied by the Ising spins. By combining the aforementioned analytical procedure with known exact solutions of the corresponding Ising archimedean lattices, various intriguing frustration effects in the mixed-spin Ising-Heisenberg planar models can be comprehensively explored, namely the disordered spin-liquid phase [12], re-entrant phase transitions with two or three consecutive critical temperatures [13, 14], intermediate plateaus in the low-temperature magnetization curves [15], the quantum reduction of the spontaneous magnetization due to a large macroscopic ground-state degeneracy [16, 17], as well as unusulal local Schottky-type peaks in temperature dependecies of the specific heat [15, 18].

Motivated by previous studies, in the present paper we will consider a further exactly solvable frustrated Ising-Heisenberg model composed of identical trigonal bipyramids arranged into regular archimedean lattices, such as hexagonal, square or triangular lattices. Without loss of generality, a comprehensive study of magnetic properties of the representative 4-coordinated (square) lattice will be proposed, in order to clarify the spontaneous ground-state spin order, critical behavior, as well as temperature evolution of the quantum spin frustration.

The paper is organized as follows. In section 2, we introduce the magnetic structure of the model and the method of its exact treatment. Exact solutions for basic thermodynamic quantities and the critical temperature of the model are also presented. Section 3 contains the analysis of the ground state and critical behavior of the hybrid model and also a discussion of temperature dependecies of the pair spin correlation functions and basic thermodynamic quantities. Finally, the most interesting findings are briefly summarized in section 4.

## 2. Model and its exact solution

We consider 2D regular $q$-coordinated lattices composed of $Nq/2$ identical inter-connected structures of the trigonal bipyramid shape. The triangular base of each trigonal bipyramid is formed by the Heisenberg spins $S = 1/2$, while other vertices are occupied by Ising spins $\sigma = 1/2$ (see figure 1). Provided that different elementary units are inter-connected via common Ising spins and that exchange interactions are realized just between the nearest spin neighbors, the spin-1/2 Ising-Heisenberg model can be viewed as 2D spin-1/2 Ising lattice whose bonds are decorated by the Heisenberg triangular clusters. From this point of view, the total Hamiltonian of the proposed quantum-classical model can be written as a sum of $Nq/2$ commuting cluster Hamiltonians $\hat{\mathcal{H}}_j$:

$$\hat{\mathcal{H}} = \sum_{j=1}^{Nq/2} \hat{\mathcal{H}}_j, \quad (1)$$

where $N$ denotes not only the total number of elementary units, but also the total number nodal lattice sites occupied by the Ising spins. The cluster Hamiltonian $\hat{\mathcal{H}}_j$ appearing in Eq. (1) involves all exchange interactions realized between spins of the $j$th elementary unit:

$$\hat{\mathcal{H}}_j = -\sum_{k=1}^{3} \left[ J_H\big(\hat{\mathbf{S}}_{j,k}\cdot\hat{\mathbf{S}}_{j,k+1}\big)_\Delta + J_I \hat{S}^z_{j,k}(\hat{\sigma}^z_j + \hat{\sigma}^z_{j+1}) \right]. \quad (2)$$

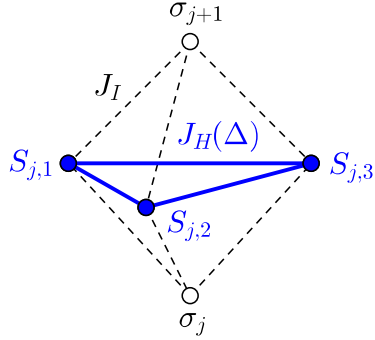

**Figure 1.** (Color online) The $j$th elementary unit of the spin-1/2 Ising-Heisenberg planar model. White (blue) circles label lattice sites occupied by the Ising (Heisenberg) spins and dashed black (solid blue) lines illustrate the Ising-type (Heisenberg-type) exchange interactions.

In above, $\big(\hat{\mathbf{S}}_{j,k} \cdot \hat{\mathbf{S}}_{j,k+1}\big)_\Delta = \Delta(\hat{S}^x_{j,k}\hat{S}^x_{j,k+1} + \hat{S}^y_{j,k}\hat{S}^y_{j,k+1}) + \hat{S}^z_{j,k}\hat{S}^z_{j,k+1}$, whereas $\hat{S}^\alpha_{j,k}$ ($\alpha \in \{x,y,z\}$) and $\hat{\sigma}^z_j$ label spatial components of the spin-1/2 operators of the Heisenberg and Ising spins, respectively, which satisfy the periodic boundary conditions $\hat{S}^\alpha_{j,4} \equiv \hat{S}^\alpha_{j,1}$ and $\hat{\sigma}^z_{Nq/2+1} \equiv \hat{\sigma}^z_1$. The parameter $J_H$ denotes the XXZ Heisenberg interaction between the nearest-neighboring Heisenberg spins, $\Delta$ is the exchange anisotropy in this interaction, and $J_I$ represents the Ising interaction between the nearest-neighboring Ising and Heisenberg spins.

Because of different cluster Hamiltonians (2) fulfill the commutation relation $[\hat{\mathcal{H}}_j, \hat{\mathcal{H}}_k] = 0$ ($j \neq k$), the summation over spin degrees of freedom of the Heisenberg spins from different trigonal bipyramid units can be realized independently of each other. Consequently, the partition function of the investigated spin-1/2 Ising-Heisenberg model can be written in the following compact form:

$$\mathcal{Z} = \sum_{\{\sigma_n\}} \prod_{j=1}^{Nq/2} \mathrm{Tr}_j \, e^{-\beta\hat{\mathcal{H}}_j}, \tag{3}$$

where $\beta = 1/(k_\mathrm{B}T)$ ($k_\mathrm{B}$ is the Boltzmann's constant and $T$ is the absolute temperature of the system), the sum $\sum_{\{\sigma_n\}}$ denotes a summation over all possible spin configurations of the Ising spins, the product $\prod_{j=1}^{Nq/2}$ runs over all trigonal bipyramid units and the symbol $\mathrm{Tr}_j$ stands for a trace over the spin degrees of freedom of the Heisenberg spins from the $j$th trigonal bipyramid. After realization of the latter trace, one obtains the effective Boltzmann's weight $w(\sigma_j, \sigma_{j+1})$, which depends merely on the Ising spins from $j$th elementary unit. It can be replaced by a simpler equivalent function by performing the generalized decoration-iteration transformation [19, 20]:

$$\begin{aligned}
w(\sigma_j, \sigma_{j+1}) &= \mathrm{Tr}_j \, e^{-\beta\hat{\mathcal{H}}_j} \\
&= 2e^{\beta J_H(4\Delta-1)/4} \cosh\left[\frac{\beta J_I}{2}(\sigma_j + \sigma_{j+1})\right] \\
&\quad + 2e^{-\beta J_H(1+2\Delta)/4} \cosh\left[\frac{\beta J_I}{2}(\sigma_j + \sigma_{j+1})\right] \\
&\quad + 2e^{3\beta J_H/4} \cosh\left[\frac{3\beta J_I}{2}(\sigma_j + \sigma_{j+1})\right] \\
&= A e^{\beta J_{eff}\sigma_j\sigma_{j+1}}.
\end{aligned} \tag{4}$$

From the physical point of view, applying decoration-iteration transformation means to replace the Heisenberg spins and all associated interactions by a new effective exchange interaction $J_{eff}$ between remaining Ising spins. The parameters $A$ and $J_{eff}$ appearing in the last line of Eq. (4) are unambiguously determined by a 'self-consistency' of the used algebraic technique:

$$A = \left(w_+ w_- w_0^2\right)^{1/4}, \tag{5}$$

$$J_{eff} = k_\mathrm{B} T \ln\left(\frac{w_+ w_-}{w_0^2}\right), \tag{6}$$

where $w_+ = w(1/2, 1/2)$, $w_- = w(-1/2, -1/2)$ and $w_0 = w(1/2, -1/2) = w(-1/2, 1/2)$.

After substituting Eq. (4) into the formula (3), one obtains the universal rigorous correspondence between the partition function $\mathcal{Z}$ of the considered spin-1/2 Ising-Heisenberg planar model and the partition function $\mathcal{Z}_I$ of the corresponding pure spin-1/2 Ising lattice given by the Hamiltonian $\mathcal{H}_I = -J_{eff} \sum_{\langle j,n \rangle}^{Nq/2} \sigma_j^z \sigma_n^z$:

$$\mathcal{Z} = A^{Nq/2} \mathcal{Z}_I. \tag{7}$$

It is worthy to note that the partition function of the spin-1/2 Ising model was rigorously calculated for several archimedean lattices [21–25]. Hence, the mapping relation (7) gives an opportunity to rigorously calculate all important thermodynamic quantities, as well as critical temperature of the studied model regardless of the lattice coordination number $q$.

*2.1. Free and internal energies, entropy, specific heat*

To be specific, by applying the standard relations of thermodynamics to Eq. (7), the Helmholtz free energy $\mathcal{F}$ of the spin-1/2 Ising-Heisenberg model defined by the Hamiltonian (1) can be expressed by means of the Helmholtz free energy $\mathcal{F}_I$ of the corresponding spin-1/2 Ising lattice with the effective nearest-neighbor coupling given by Eq. (6):

$$\mathcal{F} = -k_\mathrm{B} T \ln \mathcal{Z}_{IH} = \mathcal{F}_I - \frac{k_\mathrm{B} T N q}{8} \ln (w_+ w_- w_0). \tag{8}$$

Likewise, the internal energy $\mathcal{U}$ of the model can be connected to the known internal energy $\mathcal{U}_I$ of the corresponding Ising lattice:

$$\mathcal{U} = -\frac{\partial}{\partial \beta} \ln \mathcal{Z} = \frac{\mathcal{U}_I}{2 J_{\mathit{eff}}} \left( \frac{v_+}{w_+} + \frac{v_-}{w_-} - 2 \frac{v_0}{w_0} \right)$$
$$- \frac{Nq}{16} \left( \frac{v_+}{w_+} + \frac{v_-}{w_-} + 2 \frac{v_0}{w_0} \right). \tag{9}$$

Exact expressions for $\mathcal{F}_I$ and $\mathcal{U}_I$ can be found e.g. in [23], and the functions $v_\pm$, $v_0$, emerging in Eq. (9), are explicitly given by the formula:

$$v_\gamma = J_H(4\Delta - 1) \mathrm{e}^{\beta J_H (4\Delta - 1)/4} \cosh\left(\frac{\gamma \beta J_I}{2}\right)$$
$$+ 2\gamma J_I \mathrm{e}^{\beta J_H (4\Delta - 1)/4} \sinh\left(\frac{\gamma \beta J_I}{2}\right)$$
$$- J_H(1 + 2\Delta) \mathrm{e}^{-\beta J_H (1 + 2\Delta)/4} \cosh\left(\frac{\gamma \beta J_I}{2}\right)$$
$$+ 2\gamma J_I \mathrm{e}^{-\beta J_H (1 + 2\Delta)/4} \sinh\left(\frac{\gamma \beta J_I}{2}\right)$$
$$+ 3 J_H \mathrm{e}^{3\beta J_H / 4} \cosh\left(\frac{\gamma 3 \beta J_I}{2}\right)$$
$$+ \gamma 6 J_I \mathrm{e}^{3\beta J_H / 4} \sinh\left(\frac{\gamma 3 \beta J_I}{2}\right), \qquad \gamma \in \{\pm 1, 0\}.$$

At this calculation level, the entropy $\mathcal{S}$ and the specific heat $\mathcal{C}$ of the investigated quantum-classical model can be exactly derived by substituting the expressions (8) and (9) into the fundamental thermodynamic relations:

$$\mathcal{S} = \frac{\mathcal{U} - \mathcal{F}}{T}, \qquad \mathcal{C} = \frac{\partial \mathcal{U}}{\partial T}. \tag{10}$$

The final explicit formulas are rather extensive, but very similar to the ones for the Helmholtz free energy (8) and/or the internal energy (9), therefore we leave them to the reader.

*2.2. Magnetization and pair correlation functions*

Now, let us turn our attention to the calculation of the spontaneous magnetization and pair correlation functions of the spin-1/2 Ising-Heisenberg model. By combining the relation (7) with exact mapping theorems developed by Barry *et al.* [26–28], the following exact identities for the spontaneous magnetization $M_I$ per Ising spin and the correlation function $C_{II}^{zz}$ indicating the spin arrangement of its nearest Ising pairs in the spin-1/2 Ising-Heiseberg model can be easily acquired:

$$M_I \equiv \langle \hat{\sigma}_j^z \rangle = \langle \sigma_j^z \rangle_I = m_I, \tag{11}$$

$$C_{II}^{zz} \equiv \langle \hat{\sigma}_j^z \hat{\sigma}_{j+1}^z \rangle = \langle \sigma_j^z \sigma_{j+1}^z \rangle_I = c_I. \tag{12}$$

In above, the symbols $\langle \ldots \rangle$ and $\langle \ldots \rangle_I$ denote the standard canonical averages performed over the investigated quantum model and the corresponding isotropic Ising lattice, respectively. According to Eqs. (11) and (12), the quantities $M_I$ and $C_{II}^{zz}$ equal to the spontaneous single-site magnetization $m_I$ and the nearest-neighbor pair correlation function $c_I$ of the effective Ising model on the corresponding $q$-coordinated lattice with the nearest-neighbor spin coupling given by Eq. (6), respectively. Since exact solutions of latter two quantities are known for several regular spin-1/2 Ising lattices for several years [23, 29–32], our evaluations of the spontaneous magnetization $M_I$ and the correlation function $C_{II}^{zz}$ can be regarded as complete.

Further quantities such as the spontaneous magnetization $M_\triangle$ per XXZ Heisenberg triangular cluster and the pair correlation functions $C_\triangle^{xx}$, $C_\triangle^{zz}$, $C_{I\triangle}^{zz}$, which are helpful in specifying the spin-spin order between the Heisenberg spins and their nearest Ising neighbors, respectively, can be obtained by applying the generalized Callen-Suzuki spin identity [33–35]:

$$M_\triangle \equiv \langle \sum_{k=1}^{3} \hat{S}_{j,k}^z \rangle = \frac{B_I^+ f_+}{w_+} + \frac{B_I^- f_-}{w_-} + \frac{D_I f_0}{w_0}, \tag{13}$$

$$C_\triangle^{zz} \equiv \langle \sum_{k=1}^{3} \hat{S}_{j,k}^z \hat{S}_{j,k+1}^z \rangle = \frac{B_I^+ g_+}{2w_+} + \frac{B_I^- g_-}{2w_-} + \frac{D_I g_0}{2w_0}, \tag{14}$$

$$C_\triangle^{xx} \equiv \langle \sum_{k=1}^{3} \hat{S}_{i,j}^x \hat{S}_{j,k+1}^x \rangle = \langle \sum_{k=1}^{3} \hat{S}_{j,k}^y \hat{S}_{j,k+1}^y \rangle$$
$$= \frac{B_I^+ h_+}{w_+} + \frac{B_I^- h_-}{w_-} + \frac{D_I h_0}{w_0}, \tag{15}$$

$$C_{I\triangle}^{zz} \equiv \langle \hat{\sigma}_j^z \sum_{k=1}^{3} \hat{S}_{j,k}^z \rangle = \frac{B_I^+ f_+}{2w_+} - \frac{B_I^- f_-}{2w_-}. \tag{16}$$

The new parameters $B_I^\gamma$, $D_I$, $f_\gamma$, $g_\gamma$ and $h_\gamma$ ($\gamma = +$ or $-$), which have been introduced for the sake of brevity of Eqs. (13)–(16), are explicitly given by the formulas:

$$B_I^\pm = \frac{1}{4} \pm m_I + c_I, \quad D_I = \frac{1}{2} - 2c_I,$$

$$f_\gamma = e^{\beta J_H(4\Delta-1)/4}\left(1 + e^{-3\beta J_H \Delta/2}\right)\sinh\left(\frac{\gamma\beta J_I}{2}\right)$$
$$+ 3e^{3\beta J_H/4}\sinh\left(\frac{\gamma 3\beta J_I}{2}\right),$$

$$g_\gamma = -e^{\beta J_H(4\Delta-1)/4}\left(1 + 2e^{-3\beta J_H \Delta/2}\right)\cosh\left(\frac{\gamma\beta J_I}{2}\right)$$
$$+ 3e^{3\beta J_H/4}\cosh\left(\frac{\gamma 3\beta J_I}{2}\right),$$

$$h_\gamma = e^{\beta J_H(4\Delta-1)/4}\left(1 - e^{-3\beta J_H \Delta/2}\right)\cosh\left(\frac{\gamma\beta J_I}{2}\right).$$

*2.3. Critical temperature*

Last but not least, the rigorous criterion determining the critical temperature for the investigated spin-1/2 Ising-Heisenberg planar model follows directly from the mapping relation (7); the quantum-classical spin system may exhibits critical behavior only when its equivalent spin-1/2 Ising lattice is at a critical point. Taking into account this fact, the critical temperature of the model defined through the Hamitonian (1) can be obtained by comparing the effective nearest-neighbor coupling (6) of the corresponding spin-1/2 Ising planar lattice with its exact critical temperature. In this regard, we refer to the reference [36], where exact critical temperatures of a few spin-1/2 Ising archimedean lattices are summarized (page 95, table 7).

## 3. Results and discussion

At this stage, we can access the discussion of the most interesting numerical results. Even though the outcomes derived in the foregoing section are general in terms of the lattice coordination number $q$, as well as signs of exchange interactions $J_I$, $J_H$, we restrict our analysis to the spin-1/2 Ising-Heisenberg planar model on a representative bond-decorated square lattice ($q = 4$) with the ferromagnetic Ising interaction $J_I > 0$. The reason is that the quantum-classical model displays similar magnetic features for any $q$-coordinated lattice and the transformation $J_I \to -J_I$ causes just a trivial sign change of the Ising spin states, whereas ground-state and finite-temperature properties of the model remain qualitatively unchanged. On the contrary, a change of the ferromagnetic Heisenberg interaction $J_H > 0$ to the antiferromagnetic one $J_H < 0$ may essentially influence magnetic properties of the system, therefore we will analyze both the cases hereafter.

*3.1. Ground state*

We start by analyzing the spin arrangement of the model in the ground state. The mutual interplay between the interaction parameters $J_I$, $J_H$ and $\Delta$ results in three different spontaneously long-range ordered ferromagnetic phases, which are unambiguously characterized by the following eigenvectors, sublattice magnetization, pair correlation functions and energies:

(i) the classical ferromagnetic (CF) phase:

$$|\text{CF}\rangle = \prod_{j=1}^{2N} |\uparrow\rangle_{\sigma_j} \otimes |\uparrow\uparrow\uparrow\rangle_j,$$

$$\begin{aligned} &M_I = 1/2, \quad M_\triangle = 3/2 \quad C_{II}^{zz} = 1/4, \\ &C_\triangle^{zz} = 3/4, \quad C_{I\triangle}^{zz} = 3/4, \quad C_\triangle^{xx} = 0, \end{aligned}$$

$$E_{\text{CF}} = -\frac{3N}{2}(J_H + 2J_I); \quad (17)$$

(ii) the quantum ferromagnetic (QF) phase:

$$|\text{QF}\rangle = \prod_{j=1}^{2N} |\uparrow\rangle_{\sigma_j} \otimes \frac{1}{\sqrt{3}} (|\uparrow\uparrow\downarrow\rangle + |\uparrow\downarrow\uparrow\rangle + |\downarrow\uparrow\uparrow\rangle)_j,$$

$$\begin{aligned} &M_I = 1/2, \quad M_\triangle = 1/2, \quad C_{II}^{zz} = 1/4, \\ &C_\triangle^{zz} = -1/4, \quad C_{I\triangle}^{zz} = 1/4, \quad C_\triangle^{xx} = 1/2, \end{aligned}$$

$$E_{\text{QF}} = \frac{N}{2}(J_H - 4J_H\Delta - 2J_I); \quad (18)$$

(iii) the chiral ferromagnetic (CHF) phase:

$$|\text{CHF}\rangle = \prod_{j=1}^{2N} |\uparrow\rangle_{\sigma_j} \otimes \begin{cases} \frac{1}{\sqrt{3}} (|\uparrow\uparrow\downarrow\rangle + \omega|\uparrow\downarrow\uparrow\rangle + \omega^2|\downarrow\uparrow\uparrow\rangle)_j \\ \frac{1}{\sqrt{3}} (|\uparrow\uparrow\downarrow\rangle + \omega^2|\uparrow\downarrow\uparrow\rangle + \omega|\downarrow\uparrow\uparrow\rangle)_j \end{cases},$$

$$\begin{aligned} &M_I = 1/2, \quad M_\triangle = 1/2, \quad C_{II}^{zz} = 1/4, \\ &C_\triangle^{zz} = -1/4, \quad C_{I\triangle}^{zz} = 1/4, \quad C_\triangle^{xx} = -1/4, \end{aligned}$$

$$E_{\text{CHF}} = \frac{N}{2}(J_H + 2J_H\Delta - 2J_I). \quad (19)$$

Recall that the product $\prod_{j=1}^{2N}$ runs over mixed-spin trigonal bipyramids and $\omega$ is the cube root of unity. The single-site ket vector determines the up state of the $j$th Ising spin, while three-site ket vectors refer to spin arrangements of the adjacent (also $j$th) Heisenberg trimer.

As one can see from Eq. (17), the CF phase involves a perfect spontaneous ferromagnetic spin arrangement peculiar for pure Ising systems. However, the classical spin order is limited just to the interaction ratios $J_H/J_I > -2/(\Delta+2)$ without limiting the exchange anisotropy $\Delta$ and $J_H/J_I < 1/(\Delta - 1)$ by assuming $\Delta > 1$. If the ratio between Heisenberg and Ising interactions takes the values $J_H/J_I > 1/(\Delta - 1)$ and the anisotropy is solely of the easy-plane type ($\Delta > 1$), more interesting QF phase constitutes the ground state; all the Ising spins occupy up states, while the Heisenberg trimers are in a symmetric quantum superposition of three possible up-up-down spin states, see Eq. (18). In accordance with the spin arrangement of the Heisenberg trimers, the corresponding zero-temperature spontaneous magnetization $M_\triangle$ is reduced to the one third of the value observed in the previous CF phase. The same quantum reduction of $M_\triangle$ can be found in the CHF phase defined by Eq. (19). However, the CHF phase represents not only quantum, but also macroscopically degenerated ground state with the residual entropy per elementary unit $\mathcal{S}/(2Nk_B) = \ln 2 \approx 0.693$ due to two possible chiral degrees of freedom (left- and right-hand side) of the Heisenberg spin trimers. The observed local chirality of Heisenberg trimers is generated by a mutual competition between the ferromagnetic Ising interaction $J_I > 0$ and the antiferromagnetic Heisenberg interaction $J_H/J_I < -2/(\Delta + 2)$.

Last but not least, we can see from Eqs. (18), (19) that the QF and CHF phases are characterized by the same negative longitudinal correlation function $C_\triangle^{zz} = -1/4$. This guarantees negative values also for products of correlation functions along all three-site plaquettes forming the elementary unit of the investigated system, namely $(C_\triangle^{zz})^3$, running along the Heisenberg triangular cluster, and $C_\triangle^{zz}(C_{I\triangle}^{zz})^2$, which is realized along the plaquette of two Heisenberg spins of the same triange and their nearest Ising neighbor (see figure 1). With this in mind one may conclude that both QF and CHF phases are spontaneously frustrated in accordance to the concept of frustration developed by dos Santos and Lyra [37], which determines the frustrated state by a negative value of the product of pair correlation functions along an elementary plaquette.

## 3.2. Critical behavior

To determine the thermal stability of the spontaneous long-range ferromagnetic spin order observed in individual ground-state phases, the 3D view of the critical temperature of the studied model in dependence on the interaction ratio $J_H/J_I$ and the exchange anisotropy $\Delta$ is illustrated in figure 2. For better clarity, the global finite-temperature phase diagram is supplemented with shifted zero-temperature $J_H/J_I - \Delta$ plane showing the ground-state phase diagram of the system. Obviously, the classical spontaneous ferromagnetic spin order corresponding to the CF phase is thermally much more stable than those, which can be observed in the QF and CHF phases. To be specific, the highest critical temperature of the CF phase is equal to

$$\frac{k_B T_c^{\max}}{J_I} = \frac{3}{2\ln\left(1 + \sqrt{2} + \sqrt{2 + 2\sqrt{2}}\right)} \approx 0.981. \qquad (20)$$

As indicated in figure 2, it can be achieved just for infinitely strong ferromagnetic XXZ Heisenberg coupling $J_H \to \infty$ of the easy-axis type ($0 < \Delta < 1$). On the other hand,

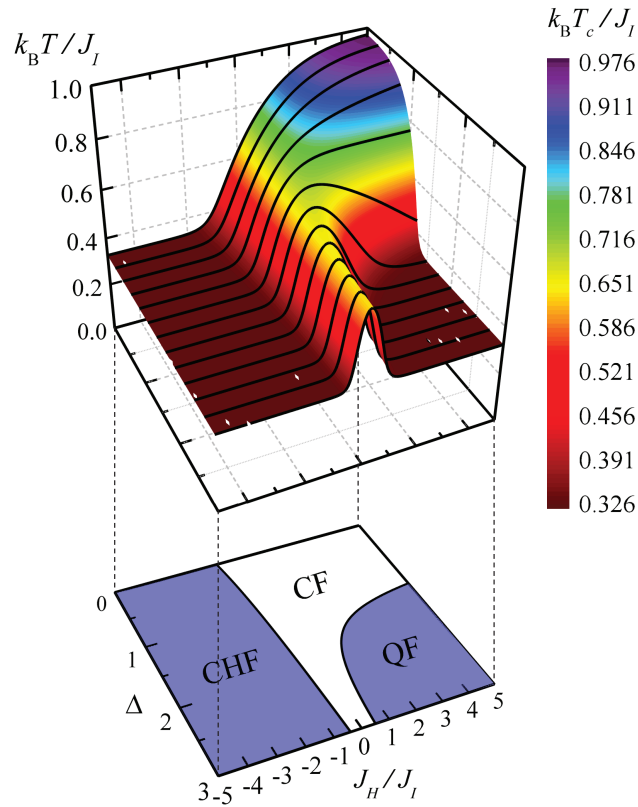

**Figure 2.** (Color online) Global finite-temperature phase diagram of the spin-1/2 Ising-Heisenberg model on bond-decorated square lattice supplemented with shifted zero-temperature $J_H/J_I - \Delta$ plane showing the ground-state phase diagram of the system.

critical temperatures of the latter two phases QF and CHF take the same constant value

$$\frac{k_\mathrm{B} T_c}{J_I} = \frac{1}{2\ln\left(1+\sqrt{2}+\sqrt{2+2\sqrt{2}}\right)} \approx 0.327 \qquad (21)$$

almost above entire their stability regions, excepting the neighborhood of the phase transitions to the CF ground state. Obviously, the critical temperature for both the quantum phases is exactly three times lower than the asymptotic value of the classical phase. The reason is a quantum reduction of the spontaneous magnetization of the Heisenberg trimers to one-third of its saturation value.

The above findings can be independently confirmed by temperature dependencies of the spontaneous magnetization $M_I$ per Ising spin and $M_\triangle$ per Heisenberg trimer, which are depicted in figure 3. As one can easily understand from this figure, the magnetization $M_\triangle$ is thermally more sensitive than $M_I$, but both they exhibit a standard steep power-law decline with the critical exponent $\beta = 1/8$ in a vicinity of the critical temperature. In accordance with the critical behavior of the model, they rapidly drop to the zero value at the critical temperature $k_\mathrm{B} T_c/J_I \approx 0.981$ if $\Delta = 0.5$ and $J_H/J_I \to \infty$, while at $k_\mathrm{B} T_c/J_I \approx 0.327$ if $\Delta = 0.5$, $J_H/J_I = -2$ and $\Delta = 2$, $J_H/J_I = 3$. Moreover, in the former case the magnetization curves start from the saturation values $M_I = 1/2$ and $M_\triangle = 3/2$ [see figure 3(a)], which provides an unambiguous evidence of the classical ferromagnetic spin order (17) in the ground state. However, in agreement with ground-state analysis, the initial value of $M_\triangle$ remains saturated just for the interaction ratios $J_H/J_I > -0.8$ if $\Delta = 0.5$ and $J_H/J_I < 1$ if $\Delta = 2$. When the opposite conditions $J_H/J_I < -0.8$, $\Delta = 0.5$ and $J_H/J_I > 1$, $\Delta = 2$ are satisfied, then the zero-temperature value of the magnetization $M_\triangle$ is reduced to one third of its saturation value due to local quantum spin fluctuations in the Heisenberg triangular clusters appearing in the CHF and QF phases [see e.g., the curves corresponding to $\Delta = 0.5$, $J_H/J_I = -0.9$ in figure 3(a) and $\Delta = 2$, $J_H/J_I = 1.5$ in figure 3(b)]. Last but not least, one may also find two special initial values of the magnetization, namely $M_\triangle = 5/6 \approx 0.833$ for the combination of interaction parameters $\Delta = 0.5$, $J_H/J_I = -0.8$ and $M_\triangle = 1$ for the combination of interaction parameters $\Delta = 1.5$, $J_H/J_I = 2$. The former value is a result of the mutual coexistence of spin arrangements observed in the CF and QF phases, while the latter one is a result of the coexistence of spin arrangements inherent to the CF and CHF phases.

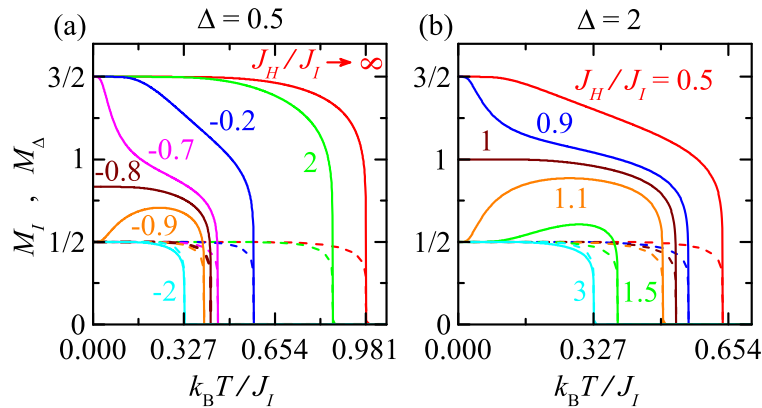

**Figure 3.** (Color online) Temperature dependencies of the spontaneous magnetization $M_I$ per Ising spin (dashed lines) and $M_\triangle$ per Heisenberg trimers (solid lines) for two representative values of the exchange anisotropy $\Delta = 0.5$ [figure (a)] and $\Delta = 2$ [figure (b)], by assuming several values of the interaction ratio $J_H/J_I$.

### 3.3. Spin frustration at finite temperatures

As mentioned in subsection 3.1, the ground-state phases QF and CHF are spontaneously frustrated in terms of dos Santos and Lyra's concept of the spin frustration [37]. Recall that the asymptotic value of the pair correlation function $C_\triangle^{zz} = -1/4$ has been used to prove a validity of the frustration conditions $(C_\triangle^{zz})^3 < 0$, $C_\triangle^{zz}(C_{I\triangle}^{zz})^2 < 0$. One can easily understand from these inequalities that the negative pair correlation function $C_\triangle^{zz}$ between $z$th components of the Heisenberg spins from the same triangle represents in fact a sufficient indicator for an incapability of the spins to simultaneously satisfy all underlying exchange interactions along each elementary cluster.

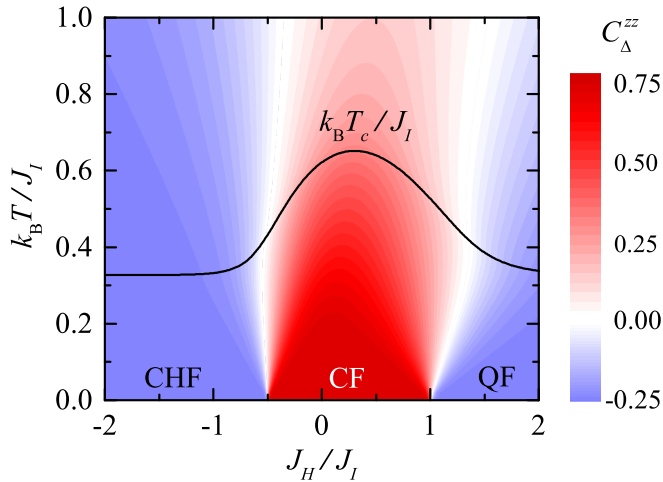

**Figure 4.** (Color online) Density plot of the pair correlation function $C_\triangle^{zz}$ in the $J_H/J_I - k_\mathrm{B}T/J_I$ plane by assuming the fixed exchange anisotropy $\Delta = 2$. Black curve shows the corresponding critical temperature $k_\mathrm{B}T_c/J_I$ of the model.

Since the frustration concept developed by dos Santos and Lyra involves the temperature, it allows one to explore the phenomenon in whole parameter space of the model, hence even at finite temperatures. The temperature effect on the spin frustration observed in QF and CHF phases is apparent from figure 4, which displays the density plot of the correlation function $C_\triangle^{zz}$ in the $J_H/J_I - k_\mathrm{B}T/J_I$ plane for the fixed exchange anisotropy $\Delta = 2$. As one can see from this figure, the spin frustration found in the Heisenberg triangular clusters persists also at finite temperatures. The phenomenon is suppressed by thermal spin fluctuations only if the QF phase constitutes the ground state, or the value of the interaction ratio $J_H/J_I$ is taken close enough to the ground-state boundaries CHF–CF, QF–CF. As a matter of fact, one, two or even three consecutive sign changes of the longitudinal pair correlation function $C_\triangle^{zz}$ can be observed above and below second-order phase transition (critical temperature) of the system in these regions.

In order to illustrate the above statements in more detail, we have plotted in figure 5 temperature dependencies of the pair correlation functions defined by Eqs. (12), (14)–(16) for several representative values of the interaction ratio $J_H/J_I$. Figures 5(a)–(c) display the situation when the frustrated CHF phase constitutes the ground state. In general, the pair correlation functions gradually decrease in their absolute values upon increasing temperature and manifest weak energy-type singularities at the critical temperature. Moreover, the negative value of the longitudinal correlation function $C_\triangle^{zz}$ demonstrates the persistent frustrated character of the spin arrangement at any temperature. The observed spin frustration of the Heisenberg spin triangles is temporarily suppressed only in a vicinity of the phase transition CHF–CF, where dominant temperature-induced arrangement of the Heisenberg trimers specific for the neighboring CF phase causes a switching of $C_\triangle^{zz}$ from negative to low positive values within some temperature range. Consequently, two frustration temperatures, at which $z$th components of the Heisenberg spins from the same triangular cluster are completely uncorrelated, can be found in this particular region. As one can see from figures 5(b), (c), the lower frustration temperature (labeled as $T_{f1}$) emerges solely below second-order phase transition, while the higher one ($T_{f2}$) is gradually moving from the ordered temperature regime $T < T_c$ to the disordered temperature regime $T > T_c$ with approaching to the ground-state boundary QF–CF. Note that spontaneous frustration of the Heisenberg spin triangles, which is characteristic for the CHF phase, is thermally evoked to a small extent even when the non-frustrated CF phase is stable in the ground state. However, this happends only if the exchange coupling $J_H$ between the Heisenberg spins is aniferromagnetic (negative) and the quantum-classical model is disordered, as we can verify in figure 5(d). Last but not least, remaining figures 5(e)–(h) show thermal evolution of the spin frustration observed in the QF phase. In accordance with general statements at the beginning of subsection, the longitudinal pair correlation function describing the relationship between $z$th components of the Heisenberg spin pairs from the same triangular cluster rapidly increases from its negative asymptotic value $C_\triangle^{zz} = -1/4$ with increasing temperature and finally turns to positive. Low positive values of $C_\triangle^{zz}$ persisting in high-temperature region clearly confirm the predicted disappearance of the spontaneous spin frustration in the Heisenberg triangular clusters above a certain (frustration) temperature. It is worth to note that the correlation function $C_\triangle^{zz}$ can change its sign not only once, but also three times upon increasing temperature. A remarkable temperature-induced re-entrance of the (non-)frustrated arrangement of the Heisenberg spins with three consecutive frustration temperatures around the critical temperature can be observed just in a narrow range of the interaction ratio $J_H/J_I \in (1.328, 1.358)$ [see figures 5(f), (g)].

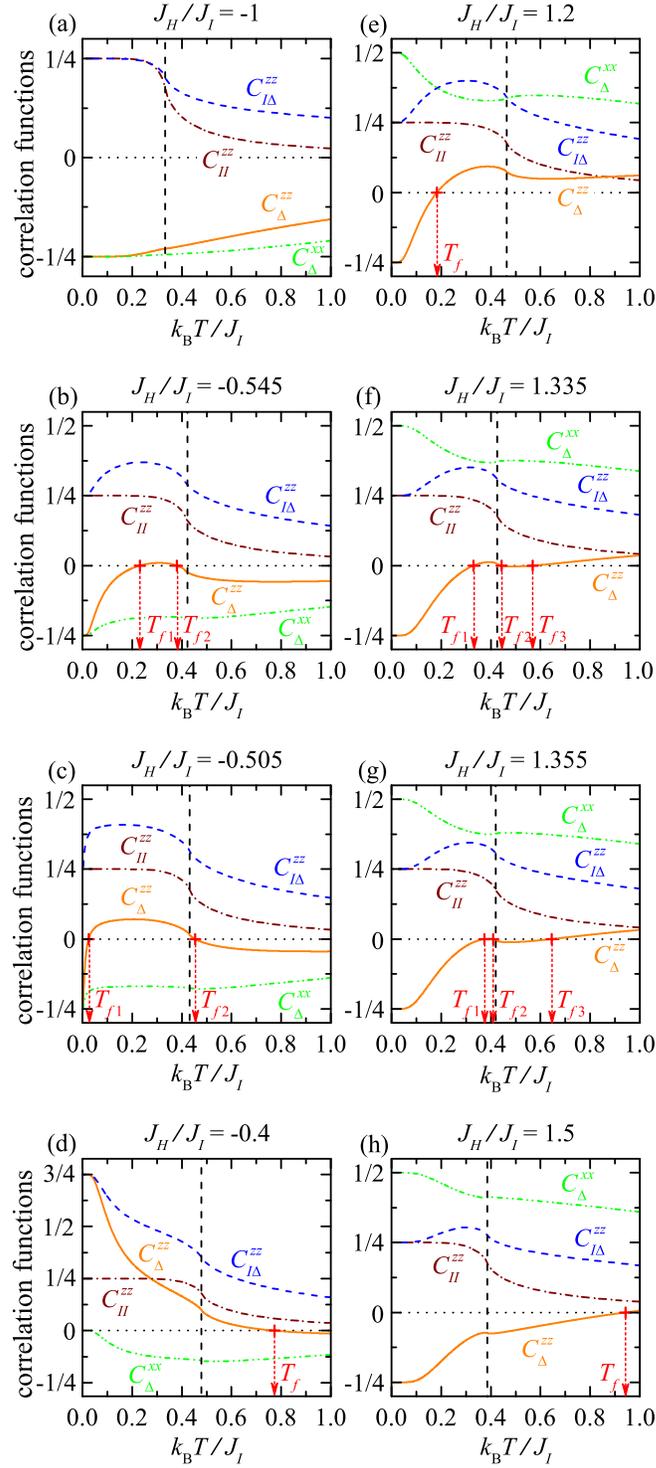

**Figure 5.** (Color online) Thermal dependencies of the pair correlation functions for the fixed exchange anisotropy $\Delta = 2$ and several representative values of the interaction ratio $J_H/J_I$. The vertical dashed lines indicate the critical temperature of the model for a considered combination of the interaction parameters. Red crosses on depicted dependencies of the correlation function $C_\triangle^{zz}$ mark positions of the frustration temperatures appearing in the temperature scale.

### 3.4. Entropy and specific heat

The macroscopic ground-state degeneracy, spin frustration as well as significant temperature variations in the sublattice magnetization and pair correlation functions may manifest themselves in unusual temperature dependencies of the entropy and specific heat. The overall picture of this issue can be gained from the density plots and temperature dependencies of the thermodynamic quantities depicted in figures 6 and 7.

It is obvious from figure 6(a) that zero-temperature entropies per elementary unit observed in individual ground-state phases generally persist within some range of low temperatures. Correspondingly, the quantity shows standard S-shaped temperature dependencies starting from one of two possible values $\mathcal{S}/(2Nk_B) = \ln 2 \approx 0.693$ or $0$ depending of whether macroscopically degenerated CHF phase or non-degenerated CF and QF phases occur in the ground state, respectively [see figure 7(a)]. Significant low-temperature variations of the entropy can be detected only in a vicinity of the ground-state phase transitions CHF–CF and CF–QF. In these particular regions, zero-temperature energies of the neighboring phases take very close values. Therefore, corresponding spin configurations become accessible at a slight temperature increase and the entropy shows clear traces of the finite value, which can be observed at the appropriate ground-state phase transition [see figures 7(c), (e)]. Further, it can be noticed from left panels of figure 7 that all the plotted $\mathcal{S}(T)$ curves also show a significant thermally induced increase with a weak energy-type singularity at the critical temperature. Its location clearly indicates that this entropy enhancement is a result of strong thermal spin fluctuations, which lead to a complete destruction of the spontanoeus magnetic order in the system. As expected, the pronounced changes in temperature dependencies of the entropy are also reflected in the corresponding temperature variations of specific heat, namely as standard logarithmic divergences nearby appropriate critical temperatures and/or low-temperature Schottky-type maxima [compare left and right panels of figure 7 and see also figure 6(b) to gain a global insight].

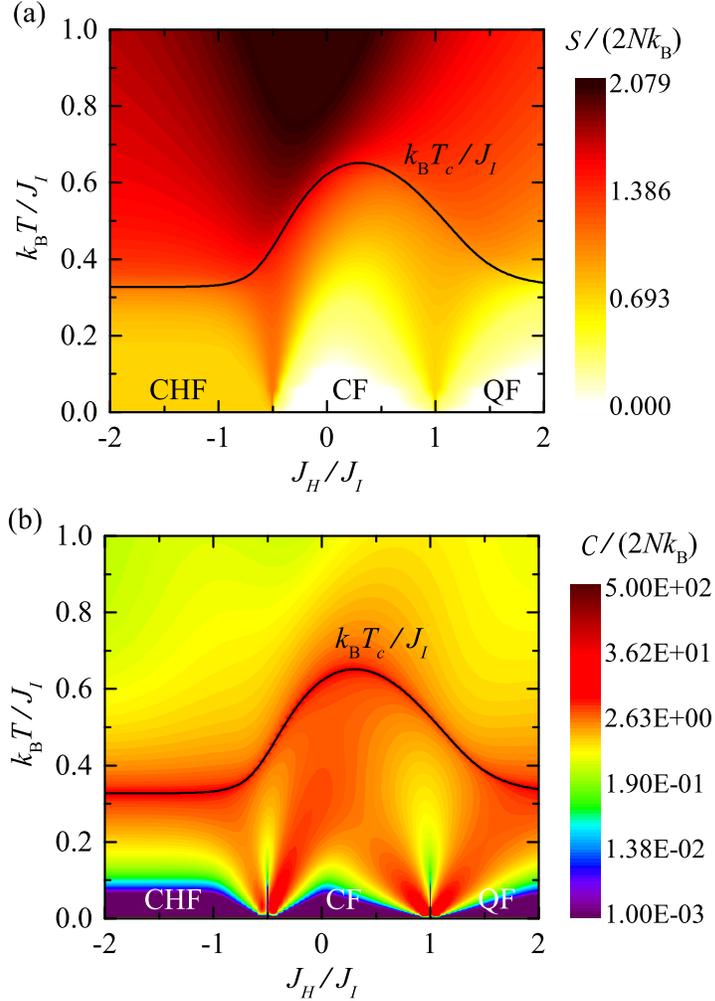

**Figure 6.** (Color online) Density plots of the entropy [figure (a)] and specific heat [figure (b)] per elementary unit in the $J_H/J_I - k_B T/J_I$ plane by assuming the fixed exchange anisotropy $\Delta = 2$. Black curve shows the corresponding critical temperature $k_B T_c/J_I$ of the investigated model.

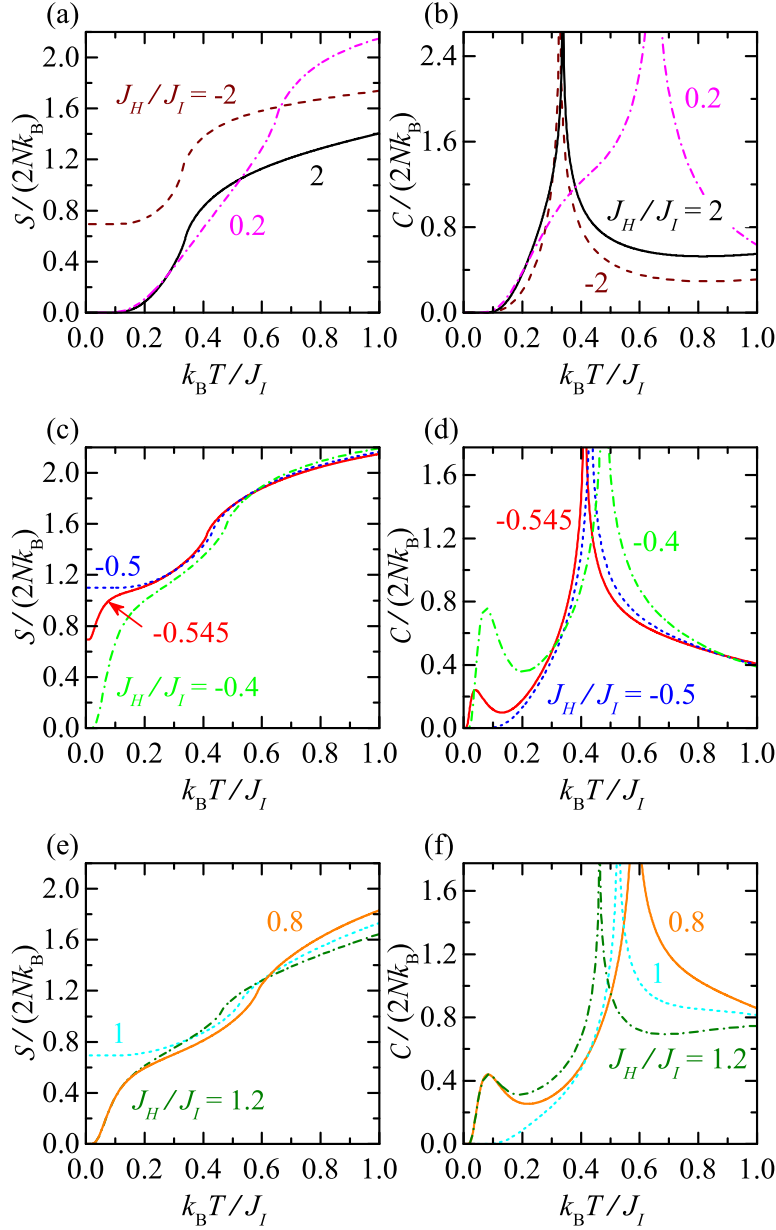

**Figure 7.** (Color online) Temperature dependencies of the entropy (left panels) and specific heat (right panels) per elementary unit for the fixed exchange anisotropy $\Delta = 2$ and several representative values of the interaction ratio $J_H/J_I$.

## 4. Conclusions

To summarize, the present article provides a survey of the most interesting exact results for the mixed spin-1/2 Ising-Heisenberg model composed of trigonal bipyramid structures that are arranged into regular archimedean lattices. Since the proposed quantum-classical model belongs to the class of the hybrid spin models on bond-decorated lattices, it was solved by means of the exact generalized decoration-iteration mapping transformation [19, 20]. Subsequently, exact results for the ground-state and finite-temperature phase diagrams, as well as several thermodynamic quantities were discussed in detail for one representative archimedean lattice - the square lattice. The main purpose of the discussion was to clarify the frustration phenomenon at zero and finite temperatures.

Our findings clearly confirm that the ground-state phase diagram of the investigated planar model contains one classical phase and two frustrated quantum phases in terms of dos Santos and Lyra's frustration concept [37]. One of the frustrated quantum phases is macroscopically degenerate due to chiral degrees of freedom of the Heisenberg spins at each triangular cluster. Interestingly, the classical long-range order is thermally much more stable than the frustrated ones, but the frustration phenomenon can persist even far above the critical temperature at which the frustrated long-range spin order completely disappears. It has been evidenced that the spin frustration can be thermally suppressed only if the ground state is constituted by the non-chiral quantum phase, or the value of the interaction ratio between Heisenberg and Ising exchange interactions is close enough to ground-state boundaries between the neighboring phases. In these particular regions, one, two, or even tree consecutive frustration temperatures, which delimit the frustrated region from the non-frustrated one, have been observed below and above critical temperature of the system. Last but not least, temperature variations of the entropy and specific heat have also been examined. It has been evidenced that both thermodymamic quantities show pronouced temperature variations not only around the critical temperature of the investigated model, but also in low-temperature regime if values of the interaction parameters are taken close enough to ground-state phase transitions.

## Acknowledgments

This work was financially supported by the grant of Slovak Research and Development Agency under the contract APVV-16-0186 and also by Ministry of Education, Science, Research and Sport of the Slovak Republic under the grands VEGA 1/0043/16, KEGA 002TUKE-4/2019.